\documentclass[aps, pra, reprint]{revtex4-1}

\usepackage{shellesc}

\usepackage{mathtools}
\usepackage{amssymb}
\usepackage{siunitx}
\usepackage{physics}

\usepackage{graphics}
\usepackage{float}
\usepackage[hidelinks, colorlinks]{hyperref}
\usepackage[]{cleveref}
\crefname{section}{sec.}{secs.}

\usepackage{xcolor}

\begin{document}

\title{The Rytov Approximation in Electron Scattering}

\author{Jonas Krehl}
\affiliation{Leibniz Institute for Solid State and Materials Research IFW Dresden, Helmholtzstr. 20, 01069 Dresden, Germany}
\email{j.krehl@ifw-dresden.de}
\author{Axel Lubk}
\affiliation{Leibniz Institute for Solid State and Materials Research IFW Dresden, Helmholtzstr. 20, 01069 Dresden, Germany}

\begin{abstract}
  In this work we introduce the Rytov approximation in the scope of high-energy electron scattering with the motivation of developing better linear models for electron scattering. Such linear models play an important role in tomography and similar reconstruction techniques. Conventional linear models, such as the Phase Grating Approximation, have reached their limits in current and forseeable applications, most importantly in achieving three-dimensional atomic resolution using electron holographic tomography. The Rytov approximation incorporates propagation effects which are the most pressing limitation of conventional models. While predominately used in the weak-scattering regime of light microscopy we show that the Rytov approximation can give reasonable results in the inherently strong-scattering regime of transmission electron microscopy.
\end{abstract}

\maketitle

\section{Introduction}

The transmission electron microscope (TEM) as an optical apparatus to measure the interaction of a specimen with electrons offers a vast array of interesting signals based on the generally strong interaction between the electrons and the sample. Coupled with the short wavelength of high-energy electrons and the optical capabilities of modern TEMs, this allows the analysis of specimen properties down to atomic resolution. The strength of the interaction mechanisms commonly makes the interpretation of the signal non-trivial, in no small part due to parasitic influences on the signal from other interactions. It is necessary to understand all of these mechanisms in order to be able to interpret the mere ``data'' the detector spews out into meaningful information about the specimen. This pertains as well to the influences of the TEM's source, optics and detector on the signal, even as we will not pay explicit consideration to these in this work. The main focus here will be the modeling of elastic electron scattering, relating properties of the specimen to the recordable signal. Due to the multitude of signals and setups TEM offers, a large range of models are needed in electron microscopy.

\subsection{Forward Models and Inverse Problems}

The elastic scattering processes involved in the beam-specimen interaction are essentially understood (e.g., \cite{Thust(2009)}, \cite{Lubk(2015)c}), so models for the elastic scattering signal for a given specimen could be constructed with an arbitrary degree of precision. These models need to be inverted for retrieving specimen information, which makes the retrieval an inverse, or rather, inversely stated, problem. Such problems appear commonly in many scientific fields, so the field of inverse problem theory \cite{louis_1989} has received, and still does, a lot of scientific effort. This theory can explain quite comprehensively, but does not eliminate, the chief problems of stating an inverse (or a sensible quasi-inverse) for a forward model and the ultimate error in the reconstructed specimen. This error encompasses the error made in the experiment (e.g. the quantization error), in the model (e.g. approximations of the interaction) and in the inversion (e.g. an approximate inversion is used or the inversion amplifies experimental noise). All these influences have to be kept in mind in an overall assessment of the reconstructed information which makes a comprehensive comparison between reconstruction methods complicated. For example, the gained accuracy of a more intricate forward model is easily lost if a more complicated inversion technique is necessary that makes the solution unstable by error amplification. This error amplification can, and usually has to, be countered by the introduction of a regularized inverse, which lowers the amplification's error contribution, while introducing an additional error term, referred to as regularization error. The art of regularization consists of choosing the strength of the regularization such to lower the overall error.

\subsection{Linear Models and their Relevance in TEM}

The theory of inverse problems becomes considerably more informative in the case of a linear forward model \cite{Hansen(1987)}. Here, an extensive theory, closely related to the spectral theory of linear operators, allows definite predictions about the solvability of inverse problems, stability of the solution, inherent error amplification as well as computationally efficient methods for the calculation of the pseudo-inverse. It is thusly quite straightforward to develop a reconstruction algorithm and to assess the accuracy and precision of that. This is critical in applications where a large set of imaging data is used to reconstruct specimen properties quantitatively.

The prime example for such an inverse problem may be electron tomography, where not only a single but a series of recordings with different object orientations is used to reconstruct the three-dimensional distribution of a specimen property \cite{Frank(2008)}. In the particular subdiscipline of electron holographic tomography the phase grating approximation (PGA) and the Lambert-Beer law have been used to reconstruct three-dimensional distributions, of electric and magnetic potentials and the transmittance respectively, down to the scale of a few nanometers (e.g., \cite{Wolf(2013)a},\cite{lubk_2014},\cite{Lubk(2014)a},\cite{Wolf(2015)}). Both of these are linear but harsh approximations of the scattering process and they produce artifacts in certain scenarios (e.g. \cite{krehl_2015}).
Recently, there has been a trend towards tomographic reconstruction schemes using non-linear approximations (e.g. \cite{vandenbroek_2012, vandenbroek_2013}) or based on prior information (e.g. \cite{chen_2015, goris_2012, jia_2014, vanaert_2011, vandenbroek_2009, vandyck_2012}). Using such complicated inversion techniques leads to severe restrictions of the sample   and uncertainty in their precision and accuracy \cite{Rez(2013)}.
To summarize: Many applications of TEM need careful treatment as inverse problems, in so far as models for the scattering process and accompanying reconstruction techniques are developed that optimize the error of the model and the reconstruction technique. The latter point suggests the preferential use of linear models for their well-developed theoretical framework. Especially in the view of advanced data reconstruction techniques such as tomography or ptychography, better models directly translate to better experimental capabilities.

In this publication we will elaborate on a particular elastic scattering approximation, which relates a measurable quantity to the electrostatic scattering potential in a linear way. The paper is organized as follows. Firstly, we introduce the basics of elastic electron scattering for kinetic energies prevalent in the TEM. Secondly, Ricatti's transformation method for solving differential equations is applied to the scattering equation. Finally, this serves as the basis for the Rytov approximation, which is the previously mentioned linear forward scattering model. In the following we discuss its main properties and compare it to frequently used approximations, such as the phase grating approximation and the Born approximation.

\section{Scalar Scattering of High-Energy Electrons}

For the purposes of imaging measurements of electron scattering in transmission electron microscopy (TEM) the electrons can be assumed to be independent from each other, since the current of the beam is small to prevent the Boersch effect in the electron source. There is no measurable influence of the electron spin so the electrons can be described as scalar particles that follow the Klein-Gordon equation (e.g., \cite{Rother(2009)a}, \cite{Spiegelberg(2015)}). The interaction with electromagnetic potentials is introduced via minimal coupling and the magnetic interaction can be neglected in high-resolution studies. With that, the starting point is a perturbed Helmholtz equation:

\begin{equation}
  \label{eq:orig_HE}
  \left(\Delta + k^2\right)\Psi(\vb{r}) = \underbrace{2 k C_e \phi(\vb{r})}_{\eqqcolon V} \Psi(\vb{r})
\end{equation}
with the wavenumber \(k\), the wave field \(\Psi\), the interaction constant \(C_e = \frac{E e}{k \hbar^2 c^2}\) and the effective potential \(V\) which is to be understood more in the sense of a interaction strength than a physical potential. Formulating it as a perturbation is not meant to imply that this problem lends itself to perturbation methods, at least not first order ones. For high energies and under neglect of inelastic scattering the potential can be assumed as static during the electron's flight. It cannot be assumed to be static during the many-electrons imaging process, in so far as different electrons scatter at different configurations of the potential, due to thermal fluctuations of the atom positions. As the intensity is averaged over this ensemble of scattering states during the measurement the experiment setup determines the influence of the thermal motion in the data, so this influence is very different between bright field images, diffraction patterns or dark field scanning TEM \cite{Rother(2009)}. We will not discuss this topic any further here. 

In TEM the modulations of the wave field due to the specimen, i.e. the right hand side of \cref{eq:orig_HE}, are small compared to the wavenumber \(k\) of the electrons. It is therefore only reasonable to separate the rapid phase oscillation of the free wave (i.e. with the wavenumber \(k\) in \(z\)-direction) by defining the wave field \(\psi\) by \(\Psi(x,y,z) = \psi(x,y,z) e^{i k z}\) yielding:
\begin{equation}
  \left(\Delta -i2k\partial_z\right)\psi(\vb{r}) = V(\vb{r}) \psi(\vb{r}) \label{eq:HE}
\end{equation}
This formulation is isomorphic to the original Helmholtz equation and makes for an equally suited starting point for further discussions. Because most of the results are more compact without the rapid phase oscillation we will use the second form if not noted otherwise.

\section{The Riccati Method}
\label{sec:riccati}

The linearity of the forward model, as desired for the solution of the inverse problem, does not strictly mean a linearity between the specimen property of interest and the measured signal. It is perfectly acceptable to bijectively, or nearly bijectively, transform both the specimen property and the signal into other quantities, if those are more suited. In the 18th century Jacopo Riccati used such a transformation to simplify linear second order differential equations; he restated the equation in terms of the logarithm of the solution and could thusly show that any linear second order differential equation can be translated to a first order quadratic equation. 
In the following we will use this transformation by defining a complex valued phase \(\Lambda(\vb{r}) = \log \psi(\vb{r})\) and evaluating the differential operators of  the inhomogeneous Helmholtz equation, yielding:
\begin{equation}
  \label{eq:preccati}
  \left(\Delta\Lambda + (\nabla \Lambda)^2 - i 2 k \partial_z\Lambda\right) e^\Lambda = V e^\Lambda
\end{equation}
We will call this phase the Rytov phase for the role it plays in the Rytov series and the Rytov approximation later on. The wave function itself \(e^\Lambda\) can be eliminated on account that it is a simple multiplicative factor (the behavior in presence of zeros in the wave function will be discussed in the next section)  on both sides with no operators acting on it:
\begin{equation}
  \label{eq:riccati}
  \left(\Delta\Lambda + (\nabla \Lambda)^2 - i 2 k \partial_z\Lambda\right) = V
\end{equation}
This equation is a quadratic first order differential equation of \(\nabla \Lambda\) because the Rytov phase \(\Lambda\) itself does not appear in the equation but only its gradient. It is, however, practical to solve this equation in terms of the Rytov phase \(\Lambda\), as this keeps the analogy to the Helmholtz equation. Especially for reasons mentioned in the next chapter it might be worthwhile to investigate the first-order form using \(\nabla \Lambda\) further.
In the absence of roots in the wave function, the solution to this Riccati equation \labelcref{eq:riccati} also solves the above equation \cref{eq:preccati} and therefore the inhomogeneous Helmholtz equation \labelcref{eq:HE} we started from.

In the end, the Riccati method replaces a multiplicative perturbation due to the scattering potential by a mere inhomogeneity at the cost of the introduction of a quadratic first order term. As a consequence of the latter, \cref{eq:riccati} is not any more amenable to analytic or numeric solutions than the perturbed Helmholtz equation we started from. As will be discussed subsequently, however, neglecting either the quadratic first order term or the linear second order term transforms \cref{eq:riccati} into known scattering approximations, namely the Rytov approximation (\cref{sec:rytov}) and the semiclassical Eikonal approximation (\cref{sec:eikonal}). Both have already shown to deliver usable results in different situations (e.g., \cite{keller_1969}, \cite{chen_1998}). For weakly scattering objects the potential itself can be treated as a perturbation and the solution can be expanded into powers of the potential giving the Rytov series \cite{born_1999, rytov_1937}.

\subsection{The Rytov Phase}

To avoid mix-ups and mathematical confusions some terms need to be used stringently in the following. A wave, wave field or wave function is a complex-valued entity as usual and consists of the amplitude (also called modulus) and the phase. The term phase always refers to the conventional phase, whereas Rytov phase refers to the complex-valued logarithm of the wave function consisting of the logarithm of the amplitude and the unwrapped phase: 
\begin{equation}
  \colorbox{green!50!black!20}{\(\displaystyle\underset{\text{Rytov~phase}\strut}{\Lambda\strut}\)} = \log\big(\colorbox{green!50!black!20}{\(\displaystyle\underset{\text{amplitude}\strut}{A\strut}\)}\big) + i \colorbox{green!50!black!20}{\(\displaystyle\underset{\text{phase}\strut}{\varphi\strut}\)}
\end{equation}

The complex logarithm in the definition of the Rytov phase is not a trivial operation and needs further discussion. Firstly, in contrast to the real logarithm it is not defined uniquely, since \(e^{i n 2 \pi}\)(\(n \in \mathbb{Z}\)) equals \num{1}. Secondly, the complex logarithm is still undefined for zeros where it diverges towards negative infinity. Both these problems appear as well in the phase-unwrapping problem of electron holography \cite{Ghiglia(1998)}, which is analogous to the complex logarithm, and consequently the same theory applies.
The first problem is well known in function theory and can be treated via the concept of analytical continuation, where the ambiguity in the value of the logarithm can be eliminated by demanding it to be analytical around every input value. These analytically continued forms of the logarithm are still non-unique with a global phase factor \(n 2 \pi\).
The wave function that solves the Helmholtz equation inside the spatial domain of a particular problem is a well-defined (but not necessarily finite) twice differentiable function. For a formulation using \(\Lambda\) to be meaningful these properties have to be expected from the Rytov phase as well. Starting from any point the continuity demand uniquely defines all values that are in a simply connected region. Thus, the pointwise ambiguity in the phase factor is eliminated and only that of the starting point remains, which means, effectively, a global ambiguity. The restriction to simply connected regions will become important together with the second problem.
At zeros of the wave function the phase is undefined, so the Rytov phase diverges towards negative infinity on the real component and is undefined in the imaginary component. In the perspective of the above derived equation \cref{eq:riccati}, which still contains the \(e^\Lambda\) terms, the elements of the equation are nonetheless well-defined by a direct correspondence to the Helmholtz equation. Only with the elimination of the \(e^\Lambda\) terms it becomes possible for the elements of the equation to be not well-defined any more. This ill-definiteness is not localized to the regions where the wave function is zero as the integral of the Rytov phase around such a region may carry an offset of multiples of \(i 2 \pi\), analogous to the winding number concept of complex analysis. This means the Rytov phase, albeit assumed continuous, cannot be represented as definite complex values any more. Other representations are easily conceivable, e.g. using \(\nabla \Lambda\) instead of \(\Lambda\) as mentioned in \cref{sec:riccati}, but this could not solve the problem of ill-definiteness at zeros of wave function. 
For now these problems cannot be resolved and have to be accepted as they are, which means that due diligence has to be paid to boundary conditions and possible zeros in the wave function when solving concrete problems.

\section{The Rytov Approximation}
\label{sec:rytov}

While it is nice to have reformulated the perturbed linear problem into a merely inhomogeneous albeit quadratic problem, quadratic differential equations are not easily solvable and the solution is not linear in the inhomogeneity However, two different approximations lend themselves by yielding manageable equations, finally realizing the aim behind using the Riccati method. Neglecting the second order term \(\Delta\Lambda\) in \cref{eq:riccati} yields a first order quadratic differential equation, the paraxial Hamilton-Jacobi equation. The ensuing semiclassical approximation is also not linear in the potential and will be discussed together with the first order Born approximation in \cref{sec:approx}. The second approximation consists of neglecting the quadratic term \((\nabla \Lambda)^2\) instead and consider the remaining equation, referred to as the Rytov approximation equation:
\begin{equation}
  \label{eq:rytov}
  \left(\Delta - i 2 k \partial_z\right)\Lambda = V
\end{equation}
That equation is virtually the same as the Helmholtz equation we started from only with a different function for which it is solved and an inhomogeneity instead of a perturbation.
The solution is straight-forward by applying the Green's operator (the operator inverse of the homogeneous part in the given boundary conditions \(G \eqqcolon \left(\Delta -i2k\partial_z\right)^{-1}\)) to the inhomogeneity and adding the free solution, yielding:

\begin{equation}
  \begin{aligned}
    \left(\Delta -i2k\partial_z\right)\Lambda_0 &= 0 \\
    \Lambda &= \Lambda_0 + G V
  \end{aligned}
\end{equation}

In consequence, the normalized Rytov phase \(\Lambda - \Lambda_0\) is linear in the potential in this approximation. Neglecting the quadratic term causes an error that is difficult to assess. With no apparent way for doing this in a general way the discussion of the error of the Rytov approximation has to be done for each concrete problem.

When considering only the homogeneous parts, i.e. the free propagation, the Rytov approximation is equivalent to the Helmholtz equation, just that the logarithm of the wave is propagated rather than the wave itself. For waves close to the unit wave these two formulations are equivalent, since \(\log(1+\partial) \approx \partial\), which makes the Rytov approximation exact for small perturbations from the unit wave.
In electron scattering this is certainly not the case. The highly localized and strong (compared to \(k\)) potential of the atoms causes a large attenuation of the wave, which does not lend itself to the treatment as a perturbation. This makes arguments and considerations like those presented in light optics \cite{slaney_1984, devaney_1981, keller_1969} inapplicable in our case but this does not mean that the Rytov approximation is inapplicable as the assumption of a small potential is an additional one.

\section{Other Approximations for Electron Scattering}
\label{sec:approx}

To put the Rytov approximation in a wider perspective and the further understand its properties, we will briefly discuss other popular approximative solutions to the Helmholtz equation, namely the first order Born approximation, the semiclassical approximation and the phase grating approximation (PGA). 

\subsection{Phase Grating Approximation}
\label{sec:PGA}

The most severe approximation considered here, the  PGA, also forms the basis of electron holographic tomography. It is largely valid under medium resolution scattering conditions, where dynamical scattering effects are suppressed by avoiding low-index zone axis conditions (e.g.,\cite{Matteucci(2002)}, \cite{Lubk(2010)a}). Here, propagation effects are neglected outright by dropping the \(\Delta \psi\)-term from the perturbed Helmholtz equation \cref{eq:HE} yielding:
\begin{equation}
  -i2k\partial_z \psi = V \psi \label{eq:PGA}
\end{equation}
The solution to this equation is a phase shift with the potential \(V\) projected along \(z\):
\begin{equation}
  \psi(z_f) = \exp\left(\frac{-i}{2 k}\int\limits_{z_i}^{z_f} \dd z V(z)\right)\; \psi(z_i)
\end{equation}
The same result can be gained by neglecting the \(\Delta\Lambda\)-term in the Rytov approximation \cref{eq:rytov}; so in the limit of neglected propagation the Rytov approximation is identical to the Helmholtz equation. In the case of elecctron holographic tomography, the neglection of the propagation has shown itself to be a serious concern at high resolution \cite{krehl_2015} and, thereby, an obstacle towards atomic resolution tomography.

\subsection{Kinematical Scattering and the Born Approximation}

In the past electron microscopy was used extensively for crystallography and took over the kinematical scattering theory of X-ray diffractometry. The kinematical theory assumes a weak interaction between the beam and the specimen, in so far as the exit wave can be assumed to be linear in the potential. This is based on the expansion of the formal solution, the Lippmann-Schwinger equation, into a Neumann series, yielding the so-called Born series:
\begin{equation}
  \psi(\vb{r}) = \sum\limits_{n=0}^{\infty} \left(\hat{G}_0 V(\vb{r})\right)^{n} \psi_0(\vb{r})
\end{equation}
Here, \(\psi_0\) is the solution of the homogeneous Helmholtz equation and \(\hat{G}_0\) is the Green's operator of the homogeneous Helmholtz equation, both of which depend on the boundary conditions of the problem. The order of the terms of this series is the power of the potential, so the zeroth order is simply the homogeneous solution, the first order the linear contribution, the second the quadratic and so on. For a small potential \(V\) this series converges to the exact solution.
In the Born approximation the series is terminated after the first order yielding the linear expression
\begin{equation}
  \psi(\vb{r}) = \psi_0(\vb{r}) + \hat{G}_0 V(\vb{r})\psi_0(\vb{r})
\end{equation}
For the higher order terms to be negligible an appropriate criterion is to assume the potential \(V\) to be very small. While that is largely the case in X-ray scattering it isn't in TEM, so the Born approximation can only serve as a point of reference for other linear models. 

\subsection{The Eikonal approximation}
\label{sec:eikonal} 

The Eikonal approximation starts very similarly to the Rytov approximation by postulating an action \(S\) by \(\Psi = \exp\left(i \frac{S}{\hbar}\right)\). This is just the Riccati method for the original Helmholtz equation \cref{eq:orig_HE} with the rapid phase oscillation, this oscillation and the scaling are the only differences between the action and the Rytov phase: \(\Lambda = i \frac{S}{\hbar} - i k z\). The wave function itself is clipped and the resulting Riccati equation is:
\begin{equation}
  \left(\frac{i}{\hbar}\Delta S - \frac{1}{\hbar^2} (\nabla S)^2 + k^2 \right) = V
\end{equation}
Due to the clipping of the wave function as in the Rytov approximation the same considerations  concerning zeros of the wave function apply. The Eikonal approximation now fully neglects the second order term \(\Delta S\) giving:
\begin{equation}
  \frac{1}{\hbar^2} (\nabla S)^2 = V - k^2
\end{equation}
An analytical solution to this stationary Hamilton-Jacobi equation may be obtained by employing the method of characteristics \cite{courant_1962}. The latter are defined by the classical equations of motion for the electron in an electric field and the phase is obtained by integrating the (reduced) action along the classical trajectories. This semiclassical or WKB approximation is applicable, if the potential varies slowly on length scales of the de Broglie wavelength of the electron. Note that more elaborate semiclassical schemes are required when classical trajectories converge (e.g. in a caustic).

\subsection{Comparison}

In scattering theory the Born approximation is also called kinematical scattering, a term which serves as contrast to dynamical scattering. This distinction is usually equated with the antonyms linear to nonlinear scattering and single to multiple scattering. With any meaningful definition of these termini these distinctions are not the same! The Born approximation is kinematical (per definition), linear (in the sense that it gives a quantity, not necessarily the wave function, that is linear in the potential) and restricted to single scattering (the resulting wave function is a superposition of the separate wave functions of each atom). The Rytov approximation is dynamical, linear and restricted to single scattering. While the Born approximation models the wave function as linear in the potential the Rytov approximation models the exponent of the wave function as linear in the potential. 
The deliberations of \cref{sec:PGA} have shown the second to be correct in a propagation-neglecting approximation. In fact, the Born series (in a initial value problem) is a Dyson series which can be expressed as an exponential where the first order is the linear order in the power series expansion of the exponential. In that line of thought the Born approximation is a weak phase approximation for each atom additionally to a single scattering approximation. The weak phase approximation is very severe and the Rytov approximation is likely to be more precise.

Two other approximations play an important role in electron microscopy: the multislice \cite{cowley_1957} and the Bloch wave approach. However, they are akin to numerical integration techniques \cite{Lubk(2015)c} and do not approximate the Helmholtz equation in the sense of the Rytov or the Born approximations. Only in some very limited cases do they give linear approximations, but their strength lies in their high accuracy and precision compared to their numerical effort. They are widely used in simulation studies for TEM experiments.

\begin{figure*}[t]
  \centering
  \renewcommand{\arraystretch}{2.0}
  \begin{tabular}{c c c}
    Helmholtz equation & paraxial approximation & ray projection\\
    \(\left(\Delta -i2k\partial_z\right)\psi = 0\) & \(\left(\partial_x^2+\partial_y^2 -i2k\partial_z\right)\psi = 0\) & \(-i2k\partial_z\psi = 0\)\\
    \(\widetilde{P}(k_x, k_y; z) = e^{i z \left(k - \sqrt{k^2-k_x^2-k_y^2}\right)}\)  & \(\widetilde{P}(k_x, k_y; z) = e^{-\frac{i z}{2 k} \left(k_x^2+k_y^2\right)}\) & \(\widetilde{P}(k_x, k_y; z) = 1\)\\
    not defined & \(P(x,y; z) = \frac{-i k}{z} e^{\frac{i k}{2 z}\left(x^2 + y^2\right)}\) & \(P(x,y; z) = \delta(x)\delta(y)\)
  \end{tabular}
  \caption{Approximations of the free propagation.}
\end{figure*}

\begin{figure*}[t]
  \centering
  %\tikzsetnextfilename{propagators}
  \includegraphics[width=.5\linewidth]{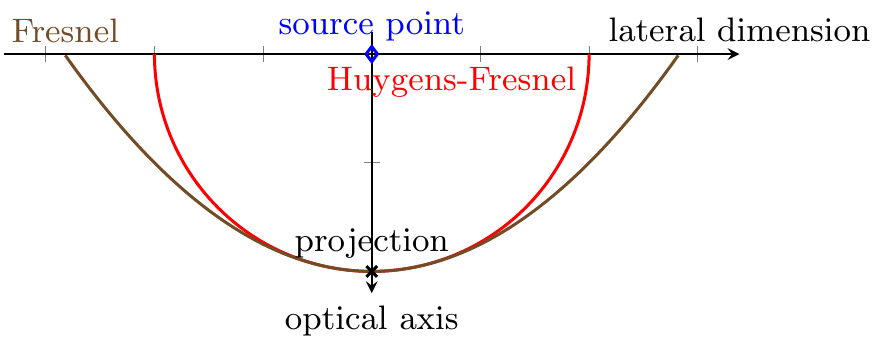}
  \caption{Approximation to the forward-only scalar wave propagation, given by the homogeneous Helmholtz equation, in the way of the Huygens-Fresnel principle by spherical waves (Huygens-Fresnel), parabolic waves (Fresnel) and corpuscles.}
\end{figure*}

\section{Free Propagation}

Of the two principal physical processes involved in scattering, i.e. free propagation and specimen interaction, that make up the transmission of an electron beam through a sample, the first one is of general interest in the comparison of scattering models.

Free propagation describes the evolution of a given wave over a given distance in the microscope in absence of any potential. The coordinate \(z\) is aligned with the optical axis of the microscope and the boundary conditions are made up of the input wave and the lateral boundaries. If we restrict ourselves to forward scattering this is an initial value problem with the evolution parameter \(z\) that has to be solved. The neglect of backwards scattering is motivated by the fact that usually only small scattering angles contribute in TEM imaging, which seems justified from practical experience but is also an ongoing field of research \cite{Spiegelberg(2015)}.

With the potential fixed at zero only the homogeneous part of the equation has to be solved, which is identical in the Helmholtz equation, the Rytov approximation and the Born approximation (see Sec. \ref{sec:approx}). It is therefore sufficient to discuss the free propagation only once, with only the physical interpretation differing between the formalisms. Here, we will use \(\psi\) and solve \(\left(\Delta -i2k\partial_z\right)\psi = 0\) as an initial value problem. 

\begin{align}
  \psi(x,y, z_f) &= P(x, y; z_f-z_i) \ast \psi(x,y, z_i)\\
  \widetilde{\psi}(k_x,k_y, z_f) &= \widetilde{P}(k_x, k_y; z_f-z_i) \widetilde{\psi}_i(k_x,k_y, z_i)
\end{align}
In the case of the propagation from \(z_i\) with the initial wave \(\psi_i\) to \(z_f\) and the sought after exit wave \(\psi_f\) the solutions are lateral convolutions or products in lateral Fourier space with a kernel \(P(x,y,z_f-zi)\).
While that can be handled quite well in different scenarios further approximations are sensible (even if it is just for computational speed), namely the paraxial approximation (i.e. Fresnel diffraction) \(\partial_z^2\psi = 0\) for small scattering angles and the ray projection approximation \(\Delta\psi = 0\) for complete neglect of any lateral propagation effects. 

They can be associated in the way of the Huygens-Fresnel principle to wave propagation by spherical waves, parabolic waves and a corpuscle. It is apparent that the parabolic wave reasonably approximates the spherical wave for small scattering angles and the projection approximation for negligible small scattering angles.

For electron scattering the paraxial approximation is usually sufficient and preferred due to its numerical simplicity. If the wave can assumed to be periodic the evaluation in Fourier space via Fast Fourier Transforms has become the predominant method.

\begin{figure*}[t]
  \centering
  %\tikzsetnextfilename{oneatomwake_comparison}
  \includegraphics[width=.95\textwidth]{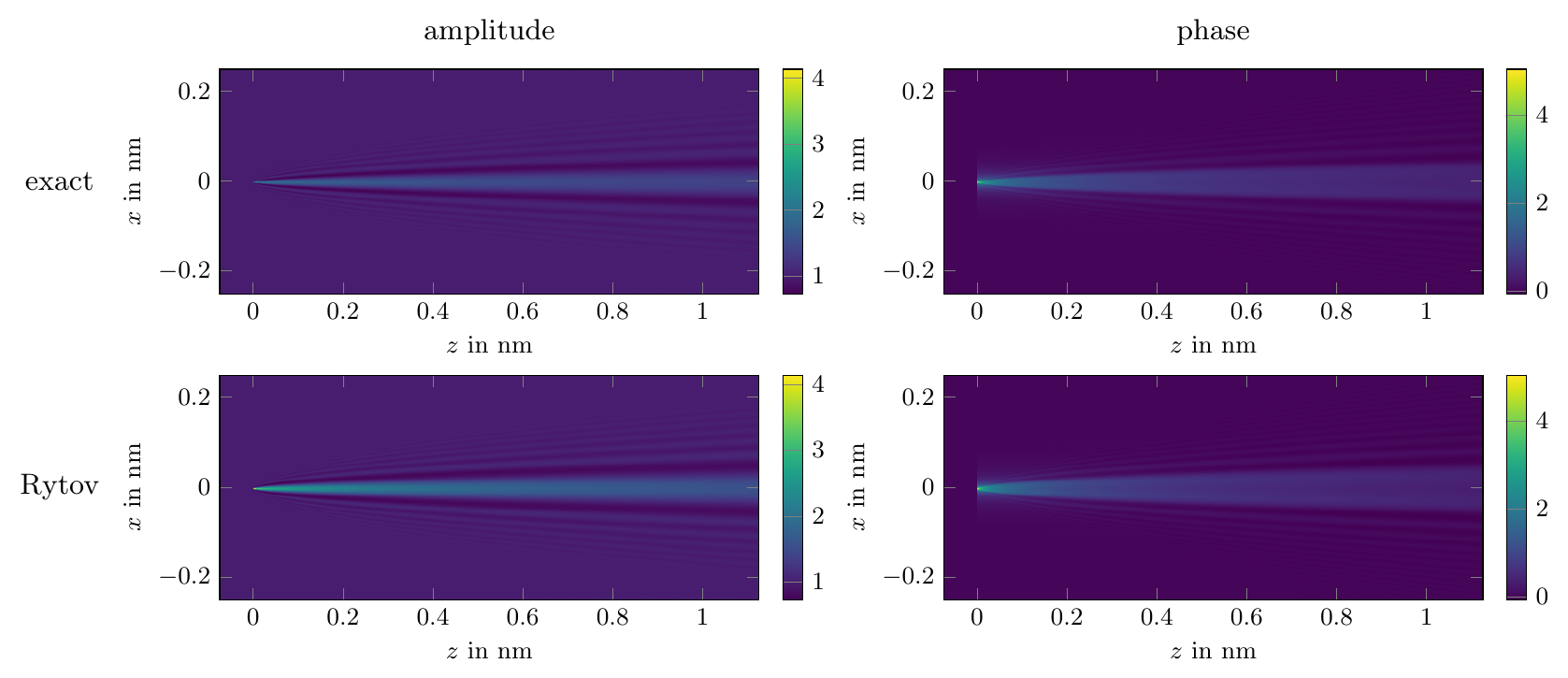}%.tikz}
  \caption{Wave field before and after a single gold atom of an \SI{200}{\keV} electron beam simulated using both a multislice approach proper and the Rytov approximation. The simulated, and therefore implicitly periodically repeated in lateral dimensions, region is four times larger than the shown region. \label{fig:oneatomwake_comparison}}
\end{figure*}

\begin{figure*}[t]
  \centering
  %\tikzsetnextfilename{oneatomwake_spectral_comparison}
  %\includegraphics[width=.95\textwidth, axisratio=4]{tikz/oneatomwake_spectral_comparison.tikz}
  \includegraphics[width=.95\textwidth]{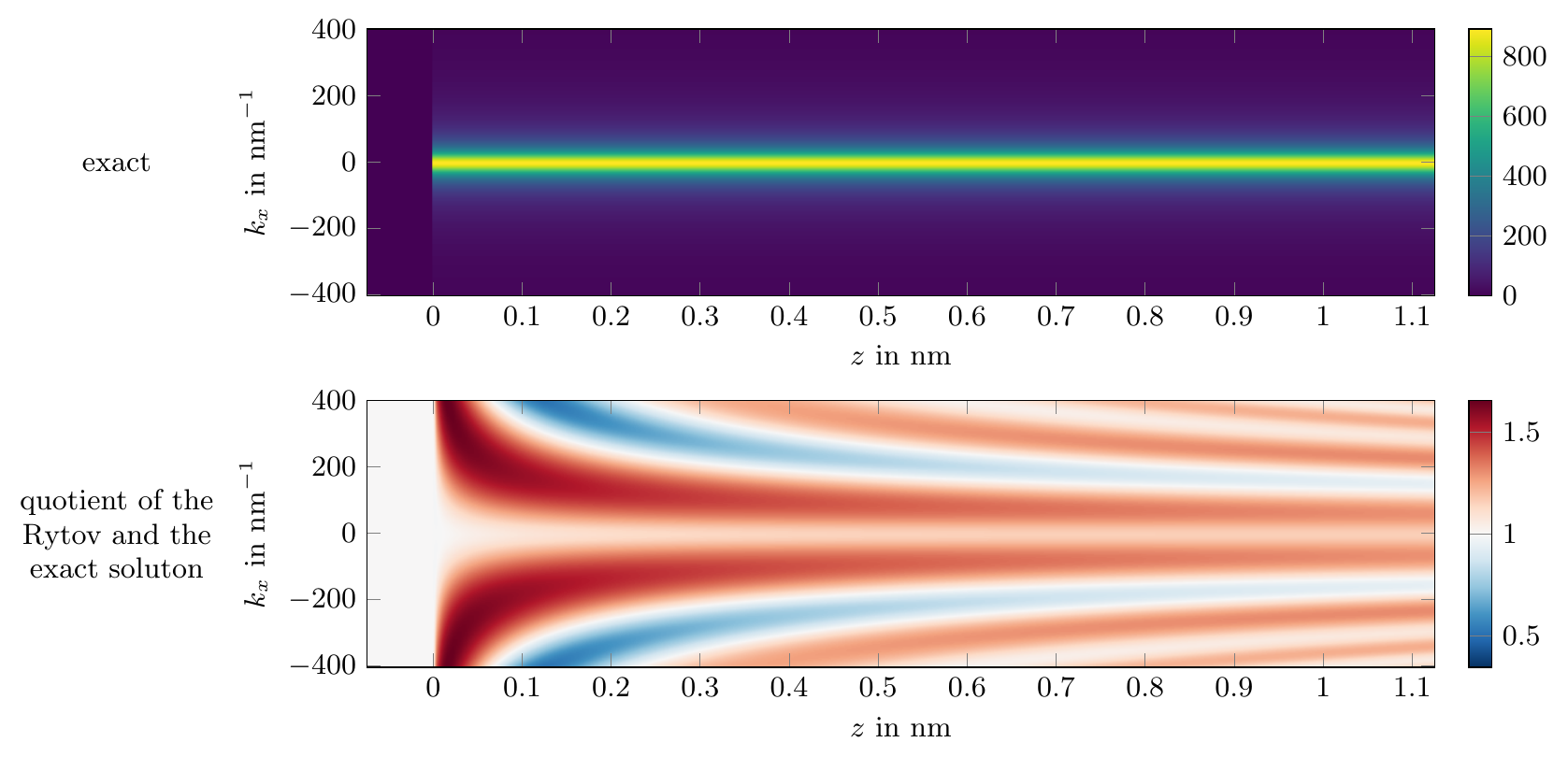}
  \caption{The absolute value of the lateral Fourier spectrum of the wave fields from \cref{fig:oneatomwake_comparison} with the Rytov approximation's relative to the exact solution.\label{fig:oneatomwake_spectral_comparison}}
\end{figure*}

\section{Wake of an Atom}

To illustrate some fundamental points about the Rytov approximation we consider the most simple exemplary system of electron scattering: the scattering of a plane wave at a single atom. 
The potential of the atom itself is collapsed along \(z\) into the plane of the center of the atom. This, let's call it, flat atom approximation stems from the multislice approach to scattering simulation and has shown to produce practical results. This does not mean that the \(z\)-dependency of the specimen is lost, but rather that the potential is region-wise concentrated into the nearest atom position. This approximation is not necessary for such a simple scenario but easier to implement in the computer.
Using a flat atom the Rytov approximation differs only in the propagation as the Fresnel propagation is here applied to the Rytov phase rather than the wave function itself. The approximations common to both solutions (elastic scattering, the flat atom, paraxial wave field, ...) are not considered here, as well as the definition problems of the Rytov phase at zeros since the incoming wave is assumed to be flat and the scattering effect is not strong enough to produce zeros. 

Inspecting the results displayed in \cref{fig:oneatomwake_comparison}, one sees that the shape of the phase is virtually identical, whereas the amplitude shows a slightly different behavior near the atom. There the amplitude in the Rytov approximation is higher by a factor of \num{\approx 1.7}. However, as the lateral spread increases with the propagation distance the values equalize.

This effect can be studied as well in the amplitude of the lateral Fourier spectrum of the wave fields (the spectrum shown in \cref{fig:oneatomwake_spectral_comparison} is of the wave field minus the incoming plane wave, in order to normalize the contrast). The amplitude for the exact solution does not change since the Fresnel propagation in Fourier space is the multiplication with a phase factor. The quickly decaying lateral behavior of the spectrum of the transmission function of a single atom is apparent and in that light it is advisable to consider the relative spectrum between the two solutions. There the difference for short propagation distances can be seen very clearly which dissipates first quickly than ever more slowly as the relative spectrum settles close to \num{1} roughly in a \SI{20}{\percent} corridor. 
The lateral region, where the spectrum assumes meaningfully large values (as seen in the upper image in \cref{fig:oneatomwake_spectral_comparison}) is quite narrow around \(k_x=0\) and of similar extent as the region of near-unity of the quotient shown in the lower image. The phase, which is not shown here, behaves well in an even larger region and is seemingly not the limiting thing here.
These arguments are, however, not exhaustive since multiple scattering scatters from large scattering angles into lower ones so the range of spatial frequencies which must be modeled correctly in the simulation becomes considerably larger than the imaged spatial frequencies. The cumulative error in multiple scattering cannot be discussed here but since the Rytov approximation is a single scattering approximation at heart it is bound to incur considerable error anyway. 

\section{Summary}

A rigorous introduction of the Rytov approximation in the context of TEM imaging is presented, at least in so far as the state of the art of electron scattering theory allows. The informative value of the ab initio discussion is limited to some points due to theoretical complexities. So, for practical information the Rytov approximation is discussed in the flat atom approximation which, in the multislice approach has shown to be an accurate model for high-resolution imaging simulations. 
The interesting relations with the Rytov approximation is however with the phase grating approximation, which neglects propagation outright, and the Born approximation, which assumes the phase shift due to the atom to be weak. In contrast to these the Rytov approximation only somewhat incorporates the propagation by applying it formally incorrect to the logarithm of the wave rather than the wave itself. For waves close to the unit wave this approximation can be shown to hold but this cannot assumed to be the case in electron scattering. With no clear route to a theoretically sound criterion apparent the Rytov approximation has to be discussed in practical terms.
 where it provides some inclusion of the propagation while maintaining linearity. While the propagation in itself is applied formally incorrect to the logarithm of the wave rather than the wave itself it models the propagation effect in some sense. 
For a simple case of a single atom it models the propagation correctly for quite a large range of lateral spatial frequencies. But problems loom for multiple scattering and very large scattering angles and will become critical for certain applications. The theoretical problems of zeros in the wave function also warrant further investigation.
In the end the Rytov approximation offers a linear approximation to electron scattering that includes the propagation effect, although only as a somewhat uncertain approximation. It does, however, model the propagation the lower spatial frequencies, which still includes those relevant for high resolution TEM, quite well. It, thusly, might prove to be an auspicious route towards atomic resolution tomography.

\begin{acknowledgements}
  This project has received funding from the European Research Council (ERC) under the European Union’s Horizon 2020 research and innovation programme (grant agreement No 715620).
\end{acknowledgements}

\bibliographystyle{apsrev4-1}
\bibliography{manuscript}

\end{document}